\documentstyle[sprocl,epsfig,wrapfig,amsmath]{article}

\begin{document}

\pagestyle{empty}

\begin{flushleft}
\large
{SAGA-HE-157-00 
\hfill January 4, 2000}  \\
\end{flushleft}
 
\vspace{1.6cm}
 
\begin{center}
 
\LARGE{{\bf Flavor asymmetry in polarized}} \\
\vspace{0.2cm}

\LARGE{{\bf proton-deuteron Drell-Yan process}} \\

\vspace{1.5cm}
 
\LARGE
{S. Kumano $^*$} \\
 
\vspace{0.4cm}
  
\LARGE
{Department of Physics}         \\
 
\LARGE
{Saga University}      \\
 
\LARGE
{Saga 840-8502, Japan} \\

\vspace{1.2cm}
 
\Large
{Talk at the RCNP-TMU Symposium on} \\

\vspace{0.2cm}

{``Spins in Nuclear and Hadronic Reactions"} \\

\vspace{0.5cm}

{Tokyo, Japan, October 26 - 28, 1999 } \\

\vspace{0.05cm}

{(talk on Oct. 28, 1999)}  \\
 
\end{center}
 
\vspace{1.3cm}

\vfill
 
\noindent
{\rule{6.0cm}{0.1mm}} \\
 
\vspace{-0.3cm}
\normalsize
\noindent
{* Email: kumanos@cc.saga-u.ac.jp.
     Information on his research is available at} \\

\vspace{-0.44cm}
\noindent
{\ \ \ http://www-hs.phys.saga-u.ac.jp.} \\

\vspace{+0.0cm}
\hfill
{\large to be published in proceedings}

\vfill\eject
\setcounter{page}{1}
\pagestyle{plain}


\title{Flavor asymmetry in polarized proton-deuteron Drell-Yan process}
\author{S. Kumano}
\address{Department of Physics, Saga University \\
         Saga 840-8502, Japan \\
         E-mail: kumanos@cc.saga-u.ac.jp}

\maketitle\abstracts
{We discuss the possibility of finding polarized antiquark flavor asymmetry
 in Drell-Yan processes. We find that the difference between polarized
 proton-proton and proton-deuteron Drell-Yan cross sections should provide
 valuable information on the polarized flavor asymmetry. Numerical results
 indicate that the asymmetry effects are conspicuous especially in the
 large-$x_F$ region. Our analysis is important for the transversity
 distributions because the flavor asymmetry cannot be found
 by inclusive lepton scattering and W-production processes.}

\section{Introduction}

Antiquark distributions used to be considered flavor symmetric
for the light antiquark distributions ($\bar u=\bar d$). However, it became
apparent that they are significantly different \cite{udbar} because of
the experimental findings of Gottfried-sum-rule violation and
proton-deuteron asymmetry in Drell-Yan experiments. Although
meson-cloud-type models seem to be the promising explanation, 
various models have been proposed to explain the experimental data.
In order to test the theoretical ideas, we need further experimental
information. In particular, polarized flavor asymmetries should provide
crucial information for determining the physics mechanism behind
the flavor asymmetric distributions. 

Longitudinally-polarized parton distributions have been investigated
mainly through the structure functions $g_1$ for the proton,
neutron (or $^3$He), and deuteron.\cite{aac} 
Semi-inclusive data were also obtained; however, they are not accurate
enough at this stage to provide any significant constraint on the polarized
distributions.\cite{my} Therefore, the antiquark distributions are assumed
to be flavor symmetric in almost all the parametrizations.\cite{aac}
We expect that the situation will become clearer by the RHIC-Spin
and other experiments. As far as the transversity is concerned,
there is no experimental information yet. Although experiments could be
done at RHIC and HERA, nobody knew how to measure the
light-antiquark flavor asymmetry because the transversity distributions
cannot be measured in the inclusive lepton scattering and W-production
experiments due to the chiral-odd property. Reference 4
proposed that the polarized proton-deuteron ($pd$) Drell-Yan process
could be used in combination with the proton-proton ($pp$) Drell-Yan
for extracting the flavor asymmetry information. 
We discuss such possibility in this talk.

In Sec. \ref{pd-dy}, relations between the polarized $pd$ Drell-Yan process
and the polarized parton distributions are introduced.
Then, we discuss how to extract the light antiquark distributions
from the polarized $pd$ Drell-Yan cross section in Sec. \ref{flavor}.
Our studies are summarized in Sec. \ref{sum}.

\section{Proton-deuteron Drell-Yan process}\label{pd-dy}

Unpolarized Drell-Yan processes have been studied as an alternative
method to lepton scattering for finding parton distributions in the nucleon
and nuclei. The unpolarized and polarized $pp$ Drell-Yan
processes have been investigated theoretically for a long time.
In addition, the unpolarized $pd$ Drell-Yan is used experimentally
for extracting the flavor-asymmetry information ($\bar u/\bar d$).
If the proton and deuteron are polarized in the $pd$ Drell-Yan
process, we could investigate a polarized version of the flavor asymmetry.
This topic is discussed in Sec. \ref{flavor}.
However, it is not straightforward to express the $pd$ cross section in
terms of structure functions. In particular, it was not clear how
the tensor structure is involved in the polarized cross section.
Reference 5 clarified this point. 

In formulating the polarized $pd$ Drell-Yan, we tried two different
methods.\cite{hk} The first one uses the Jacobi-Wick helicity formalism
with the spin-density matrices. The essential difference from the $pp$
reaction is that there exist rank-two tensors due to the spin-1 nature
of the deuteron. The conditions of Hermiticity, parity conservation,
and time-reversal invariance are imposed on possible structure
functions. Then, we found that there are 108 structure functions
in general. If they are integrated over the virtual-photon transverse
momentum $\vec Q_T$, there exist only 22 structure functions.
In the second method, the hadron tensor is expressed in terms
of possible combinations
of momentum and spin vectors by imposing the same three conditions.
Only the limiting case $Q_T\rightarrow 0$ is considered in this
analysis, and we also obtained the same 22 structure functions. 
These finite functions should be physically significant ones
which could be investigated by the polarized $pd$ Drell-Yan process.

Considering the present situation on the proton spin physics,
we think that the 22 functions are still too many to be investigated
seriously. Furthermore, the physics meaning of these functions,
particularly the new ones which do not exist in the $pp$ reaction,
is not clear. Therefore, the $pd$ Drell-Yan was also analyzed
in a parton model.\cite{hk} The hadron tensor is expressed
by correlation functions for the process
$q+\bar q\rightarrow \ell^+ +\ell^-$. Then, the correlation
functions are expanded in terms of the sixteen $4\times 4$ matrices:
${\bf 1},\, \gamma_5,\, \gamma^\mu,\, \gamma^\mu \gamma_5,\, 
    \sigma^{\mu \nu} \gamma_5$ and kinematical Lorentz vectors
and pseudovectors. 
Then, we found finite structure functions in the parton model.
There is a new polarization asymmetry $A_{UQ_0}$
with the unpolarized proton and the tensor-polarized deuteron.
It is given by the parton distributions in the proton and deuteron as
\begin{equation}
A_{UQ_0}  =  \frac{\sum_a e_a^2 \, 
                  \left[ \, f_1(x_1) \, \bar b_1(x_2)
                          + \bar f_1(x_1) \, b_1(x_2) \, \right] }
                {\sum_a e_a^2 \, 
                  \left[ \, f_1(x_1) \, \bar f_1(x_2)
                          + \bar f_1(x_1) \, f_1(x_2) \, \right] }
\, ,
\label{eqn:auq0}
\end{equation}
where $f_1(x)$ and $\bar f_1(x)$ are unpolarized quark and
antiquark distributions, and $b_1(x)$ and $\bar b_1(x)$ 
are tensor-polarized distributions. The $b_1$ structure function
is known in lepton scattering;\,\cite{b1} however, the Drell-Yan
process provides important information on the antiquark
tensor polarization $\bar b_1$.
However, this topic is no more discussed in the following because
it is not the major purpose of this paper to investigate the tensor
structure. We refer the reader to Ref. 5 for more details.
In the following, we discuss the double longitudinal and
transverse spin asymmetries in connection with the flavor asymmetry. 

First, according to the general formalism,\cite{hk} the difference
between the longitudinally-polarized $pd$ cross sections is given by
\begin{equation}
\Delta \sigma_{pd}  = \sigma(\uparrow_L , -1_L) 
                       - \sigma(\uparrow_L , +1_L) 
            \propto - \frac{1}{4} \, \left[ 2 \, V_{0,0}^{LL} 
                          + (\frac{1}{3}-cos^2 \theta ) \, 
                            V_{2,0}^{LL} \right]
\ ,
\end{equation}
where $\sigma(pol_p,pol_d)$ indicates the cross section
with the proton polarization $pol_p$ and the deuteron one $pol_d$.
The longitudinally polarized structure functions $V_{0,0}^{LL}$
and $V_{2,0}^{LL}$ are defined in Ref. 5.
The $\theta$ is the polar angle of the lepton $\ell^+$.
Then, the structure functions are related to the polarized
parton distributions in the parton-model analysis.\cite{hk}
The $\vec Q_T$-integrated results indicate
\begin{equation}
\Delta \sigma_{pd} \propto \sum_a e_a^2 \, 
             \left[ \, \Delta q_a(x_1) \, \Delta \bar q_a^{\, d}(x_2)
           + \Delta \bar q_a(x_1) \, \Delta q_a^d(x_2) \, \right]
\ ,
\label{eqn:l-pd}
\end{equation}
where $\Delta q_a^{\, d}$ and $\Delta \bar q_a^d$ are the 
longitudinally-polarized quark and antiquark distributions
in the deuteron.

In the transverse-polarization asymmetry, the situation is
more complicated in the sense that four structure functions
($V_{0,0}^{TT}$, $V_{2,0}^{TT}$, $U_{2,2}^{TT}$, and $U_{2,1}^{UT}$)
contribute. However, it becomes a simple expression if the parton
model is used by neglecting higher-twist contributions:
\begin{align}
\Delta_T \sigma_{pd} & =  \sigma(\phi_p=0,\phi_d=0)
                      - \sigma(\phi_p=0,\phi_d=\pi)
\nonumber \\
 &  \propto \sum_a e_a^2 \, 
    \left[ \, \Delta_T q_a(x_1) \, \Delta_T \bar q_a^{\, d}(x_2)
          + \Delta_T \bar q_a(x_1) \, \Delta_T q_a^d(x_2) \, \right]
\ ,
\label{eqn:t-pd}
\end{align}
where $\Delta_T q$ and $\Delta_T \bar q$ are quark and antiquark
transversity distributions,
and $\phi$ is the azimuthal angle of a polarization vector.
In this way, we found that the cross-section difference is
written in terms of the longitudinally-polarized and transversity
distributions.

\section{Light-antiquark flavor asymmetry}\label{flavor}

Because the expressions of Eqs. (\ref{eqn:l-pd}) and (\ref{eqn:t-pd})
are the same as the unpolarized one if the polarized distributions
are replaced by the unpolarized ones, the polarized flavor asymmetries 
could be extracted from the polarized $pp$ and $pd$ Drell-Yan cross
sections as it has been investigated in the unpolarized case.
In order to discuss the $pp$ and $pd$ cross sections in connection with
the flavor asymmetry, we define the ratio
\small
\begin{equation}
R_{pd}   \equiv \frac{     \Delta_{(T)} \sigma_{pd}}
                   {2 \, \Delta_{(T)} \sigma_{pp}}
         =     \frac{ \sum_a e_a^2 \, 
    \left[ \, \Delta_{(T)} q_a(x_1) \, 
              \Delta_{(T)} \bar q_a^{\, d}(x_2)
            + \Delta_{(T)} \bar q_a(x_1) \, 
              \Delta_{(T)} q_a^d(x_2) \, \right] }
              { 2 \, \sum_a e_a^2 \, 
    \left[ \, \Delta_{(T)} q_a(x_1) \, 
              \Delta_{(T)} \bar q_a(x_2)
            + \Delta_{(T)} \bar q_a(x_1) \, 
              \Delta_{(T)} q_a(x_2) \, \right] }
\ ,
\label{eqn:ratio1}
\end{equation}
\normalsize
where $\Delta_{(T)}=\Delta$ or $\Delta_T$ depending on
the longitudinal or transverse case.
There is another issue in calculating the cross sections because
of nuclear corrections. However, they are ignored in the following
discussions since they are not expected to be the essential part.
If experimental data are taken in future, such corrections should
be taken into account properly.

\begin{wrapfigure}{r}{0.46\textwidth}
   \vspace{-0.0cm}
   \begin{center}
       \epsfig{file=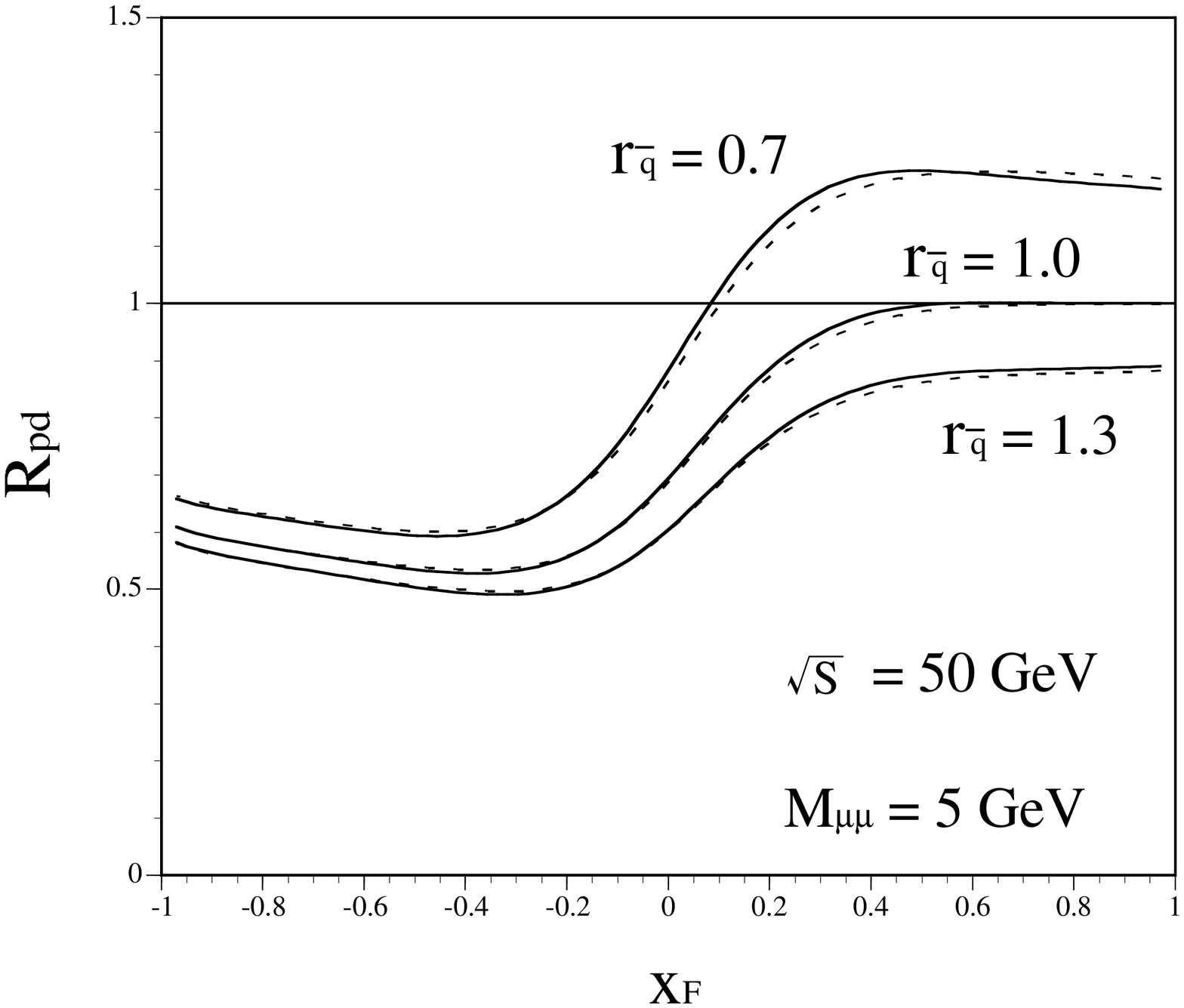,width=6.0cm}
   \end{center}
   \vspace{-0.7cm}
       \caption{\footnotesize
          Cross-section ratio $R_{pd}$ (from Ref. 4).}
       \label{fig:rpd}
\end{wrapfigure}
We show our numerical-analysis results for the ratio $R_{pd}$
in Fig. \ref{fig:rpd}. The parton distributions are taken from
Ref. 7 at $Q^2$=1 GeV$^2$, where flavor asymmetry ratio is introduced
as $r_{\bar q} \equiv \Delta_{(T)} \bar u/ \Delta_{(T)} \bar d=$0.7,
1.0, or 1.3. Then, the distributions are evolved
to $Q^2=M_{\mu\mu}$=25 GeV$^2$ by the leading-order DGLAP evolution
equations. The center-of-mass energy is taken
as $\sqrt{s}$=50 GeV with a fixed target experiment in mind.
In Fig. \ref{fig:rpd}, the solid and dashed curves are longitudinally-
and transversely-polarized ratios, respectively. Because the transversity
distributions are assumed to be the same as the corresponding
longitudinally-polarized ones at $Q^2$=1 GeV$^2$,
the transverse ratios are almost the same as the longitudinal ones.
There are large differences between the curves
for $r_{\bar q}$=0.7, 1.0, and 1.3, so that it
should be possible to extract the longitudinally-polarized and
transversity flavor asymmetries from the ratios $R_{pd}^{(L)}$ and
$R_{pd}^{(T)}$. The differences are especially large in the large-$x_F$
region with the following reason.
If two extreme limits ($x_F =x_1-x_2 \rightarrow \pm 1$) are taken in
Eq. (\ref{fig:rpd}) with the assumption 
$\Delta_{(T)} u_v (x \rightarrow 1) \gg 
 \Delta_{(T)} d_v (x \rightarrow 1)$, 
the ratio becomes
\begin{align}
R_{pd} (x_F\rightarrow +1) & =  \frac{1}{2} \, \left [ \, 1 
                 + \frac{\Delta_{(T)} \bar d (x_2)}
                        {\Delta_{(T)} \bar u (x_2)} 
                    \, \right ]_{x_2\rightarrow 0}
\, , 
\label{eqn:rpd1} 
\\
R_{pd} (x_F\rightarrow -1) & = 
     \frac{1}{2} \, \left [ \, 1 
                 + \frac{\Delta_{(T)} \bar d (x_1)}
                   {4 \, \Delta_{(T)} \bar u (x_1)} 
                    \, \right ]_{x_1\rightarrow 0}
\, .
\label{eqn:rpd2}
\end{align}
These equations suggest that the flavor-asymmetric distribution
$\Delta_{(T)} \bar u - \Delta_{(T)} \bar d$ can be extracted
by finding the deviation from 1 at $x_F\rightarrow +1$ or
from 5/8 at $x_F\rightarrow -1$. However, $R_{pd}$ should be
more sensitive to the flavor asymmetry at large $x_F$
due to the factor of 1/4 in Eq. (\ref{eqn:rpd2})
in comparison with Eq. (\ref{eqn:rpd1}).

\begin{wrapfigure}{r}{0.46\textwidth}
   \vspace{-0.0cm}
   \begin{center}
       \epsfig{file=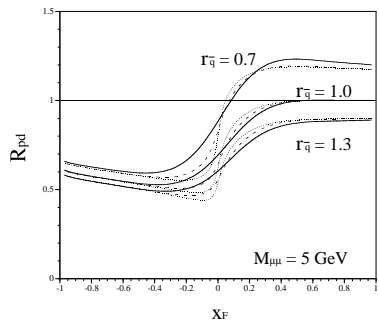,width=6.0cm}
   \end{center}
   \vspace{-0.7cm}
       \caption{\footnotesize
          $\sqrt{s}$ dependence (from Ref. 4).}
       \label{fig:s}
\end{wrapfigure}
Next, we show the numerical results in Fig. \ref{fig:s} for larger energies
$\sqrt{s}$=200 and 500 GeV by considering a collider option.
The solid, dashed, dotted curves are for $\sqrt{s}$=50, 200, and 500 GeV,
respectively. Although there are large variations in the medium-$x_F$
region, the ratios in the large- and small-$x_F$ regions stay the same.
Furthermore, we studied the parametrization-model dependence,\cite{km}
and the results indicate that the large- and small-$x_F$ ratios are again
rather independent of the parametrization. Therefore, these regions are 
appropriate for investigating the flavor asymmetry. In addition,
the variations in the medium-$x$ region indicate that 
the details of the polarized parton distributions could be
investigated in this region.

In this way, we find that it is possible to extract both
$\Delta \bar u / \Delta  \bar d$ and $\Delta_T \bar u / \Delta_T \bar d$
from the cross-section ratios, particularly in the large-$x_F$ region.
Our suggestion should be important for the transversity
because it cannot be measured in the inclusive lepton scattering
and W-production processes.

At this stage, no actual experimental measurement is planned.
However, there are certain possibilities at Fermilab, HERA-N, and JHF
for measuring the polarized $pd$ Drell-Yan cross sections
by using a fixed deuteron target. In addition, the polarized deuteron
could be accelerated in principle at RHIC. However, it is not easy
to attain the longitudinal polarization due to the small magnetic
moment unless someone has a smart idea for the longitudinal
polarization.\cite{long-d}
In any case, we should be able to investigate at least
the transverse part at RHIC. As far as the tensor polarization 
is concerned in Eq. (\ref{eqn:auq0}), we may combine the transverse
cross sections with the unpolarized one for getting the tensor-polarized
cross section. Because there are a variety of interesting topics on
polarized deuteron reactions, we hope that experimental possibilities
are seriously studied.

\section{Summary}\label{sum}

First, we briefly discussed the general framework of the polarized
$pd$ Drell-Yan process. Then, we explained the relation between
the flavor-asymmetry ratio
$\Delta_{(T)} \bar u / \Delta_{(T)} \bar d$ and the Drell-Yan
cross-section ratio
$\Delta_{(T)} \sigma_{pd}/[2 \, \Delta_{(T)} \sigma_{pp}]$.
Our numerical analysis suggested that the polarized flavor asymmetry
could be extracted from the $pd$ and $pp$ cross-section measurements
particularly in the large-$x_F$ region. At this stage, this is the
only proposal for extracting the transversity asymmetry
$\Delta_T \bar u / \Delta_T \bar d$ due to the chiral-odd property.


\section*{Acknowledgments}
S.K. was partly supported by the Grant-in-Aid for Scientific Research
from the Japanese Ministry of Education, Science, and Culture under
the contract number 10640277.


\end{document}